\documentclass[aps,prl,twocolumn,showpacs,preprintnumbers,amsmath,amssymb,10pt]{revtex4-1} 
\usepackage{amsmath,graphicx,subfig,xcolor,verbatim}
\usepackage{diagbox}

\newcommand{\be}{\begin{equation}}
\newcommand{\ee}{\end{equation}}
\newcommand{\bea}{\begin{eqnarray}}
\newcommand{\eea}{\end{eqnarray}}
\newcommand\restr[2]{{\left.\kern-\nulldelimiterspace#1\vphantom{\big|}\right|_{#2}}}

\newcommand{\ud}{\mathrm d}

\def\eps{\epsilon}
\newcommand{\beq}{\begin{equation}} 
\newcommand{\eeq}{\end{equation}}

\def\geq{\geqslant}
\def\leq{\leqslant}

\def\eps{\epsilon}

\begin{document}
\title{Taxonomy of coupled minimal models from finite groups}
\author{Ant\'onio Antunes, No\'e Suchel} 
\affiliation{
 Laboratoire de Physique de l'\'Ecole Normale Sup\'erieure, Universit\'e  PSL, CNRS, Sorbonne Universit\'e, Universit\'e  Paris Cit\'e, 24 rue Lhomond, F-75005 Paris, France}

\begin{abstract}
Fixed points of $N$ coupled Virasoro minimal models have recently been argued to provide large classes of compact unitary CFTs with $c>1$ and only Virasoro chiral symmetry. In this paper, we vastly increase the set of such potential irrational fixed points by considering couplings that break the maximal $G=S_N$ symmetry into various subgroups $H\subset G$. We rigorously classify all the fixed points with $N=4,5$ and do an extensive search for solutions of the beta function equations with $N\geq6$. In particular, we find non-trivial fixed points with $H=\mathbb{Z}_{N-1} \rtimes \mathbb{Z}_2, \, S_{M}\times S_{N-M}$ and rigorously prove that real fixed points with $H=(S_{N/2}\times S_{N/2})\rtimes \mathbb{Z}_2$ exist for all even $N\geq6$. We also identify fixed points with finite Lie-type symmetry $H=\rm{PSL}_2(N)\subset S_N$ where $N=7,11,13$ and uncover a non-unitary fixed point with $H=M_{22}\subset S_{22}$, a sporadic Mathieu group. Along the way, we encounter conformal manifolds at leading order in perturbation theory which we resolve at sub-leading order. 
\end{abstract}
\maketitle
\nopagebreak

{\bf Introduction.}
The space of conformal field theories (CFTs) is vast and diverse but full of lampposts. In four spacetime dimensions, our understanding of superconformal field theories \cite{Argyres:2022mnu} far outshines the superficial knowledge we possess on Caswell-Banks-Zaks fixed points \cite{Caswell:1974gg,Banks:1981nn,DiPietro:2020jne}. In three dimensions, the glow of $O(N)$ symmetric scalar theories \cite{Henriksson:2022rnm} offuscates a broad class of scalar CFTs with discrete symmetries relevant in real world ferromagnets \cite{Chester:2020iyt} and crystallographic phase transitions \cite{ToledanoBook,Kousvos:2018rhl,Kousvos:2025ext}. In two dimensions, exactly solvable rational CFTs with a discrete spectrum take the spotlight \cite{Belavin:1984vu}, leaving their irrational cousins in the dark \cite{Yin:2017yyn}.

In this paper, we try to open the curtains and get a glimpse of a shadowy corner in the space of 2d CFTs: IR fixed points of $N$ coupled Virasoro minimal models preserving a discrete global symmetry $H\subset G=S_N$. Coupled minimal models have been studied since the 90's \cite{djlp98} but have been recently revived as a tool to construct compact, unitary CFTs with $c>1$ and only Virasoro symmetry (which are in particular irrational) \cite{Antunes:2022vtb,Antunes:2024mfb,Antunes:2025huk,Bootstrap}. Some evidence has been given for this property in the form of anomalous dimensions for all additional current candidates up to spin $J\leq10$ when $N\geq5$. Rather than improving on the depth of these results, we will focus on their breadth \cite{Erramillitalk}. We ask: \textit{How general can the couplings between copies be while retaining the existence of non-trivial IR fixed points?}

This idea is of course far from new and has been thoroughly explored in the context of the more traditional Wilson-Fisher fixed points \cite{Wilson:1971dc}, where a Lagrangian model constructed out of free fields is studied slightly below its upper critical dimensional $d_{\rm{uc}}$ by expanding in $\epsilon=d_{\rm{uc}}-d$. The minimal setup is to take quartic scalar models with the action
\begin{equation}
\label{WFaction}
    S= \int d^{4-\epsilon}x\,  \left( \frac{1}{2}  (\partial \phi_i)^2  + \lambda_{ijkl} \phi_i\phi_j\phi_k\phi_l\right)\,,
\end{equation}
where $i=1,\dots,N$, and the fixed points are weakly coupled since $\lambda^*_{ijkl}\propto\epsilon$. Such fixed points have been rigorously classified for $N\leq5$ \cite{Toledano:1985prb,Rong:2023xhz}, and several families can be shown to exist for general (sufficiently large) $N$ \cite{Osborn:2017ucf,Rychkov:2018vya,Osborn:2020cnf,Hogervorst:2020gtc}. Apart from the maximally symmetric $O(N)$ models, there are hyper-cubic fixed points with $H=S_N \rtimes (\mathbb{Z}_2)^N$, hyper-tetrahedral fixed points $H=S_{N+1}\times\mathbb{Z}_2$ and bi-conical fixed points with $O(N_1)\times O(N_2)$ symmetry among others \cite{Henriksson:2020fqi}. This approach has also been extended to scalar-fermion models \cite{Pannell:2023tzc,Pannell:2025ajf,Mitchell:2025oao}, as well as line defects \cite{Pannell:2023pwz}, and more general surface defects and interfaces \cite{Harribey:2024gjn,Anataichuk:2025zoq,Bartlett-Tisdall:2025kcx}.

To generalize this discussion to coupled minimal models, where the spacetime dimension is fixed at $d=2$, an alternative parameter allowing for an $\epsilon$ expansion is needed. The idea of \cite{Antunes:2022vtb} is to adopt Zamolodchikov's  $\epsilon=1/m$ expansion \cite{Zamolodchikov:1987ti}, which is well-known to be useful in describing the RG flow between adjacent unitary minimal models $\mathcal{M}_m\to\mathcal{M}_{m-1}$, to the case of $N\geq 4$ minimal models by introducing a deformation that couples the replicas four at a time while remaining weakly relevant, in full analogy with \eqref{WFaction}. The original setup of \cite{Antunes:2022vtb} preserved an $S_N$ symmetry and had only two symmetry preserving couplings. In the remainder of this paper, we will explore the effect of breaking the symmetry to $H\subset G$: We will find a rich space of solutions to the fixed point equations, and therefore a large set of candidate compact unitary CFTs with $c>1$ and minimal chiral symmetry. Some of the richness of this space will turn out to be inherited from another famously rich space: the space of finite groups.  

{\bf The setup.}
We consider $N$ coupled unitary diagonal Virasoro minimal models $\mathcal{M}_m$, each with central charge
\begin{equation}
    c_m=1-\frac{6}{m(m+1)}\,,
\end{equation}
in the limit $m\to \infty$, where the Kac table Virasoro primaries $\phi_{(r,s)}$ have scaling dimension
\begin{equation}
    \Delta_{(r,s)}=2h_{(r,s)}= \frac{(r-s)^2}{2}+ \frac{r^2-s^2}{2m}+ O(m^{-2})\,.
\end{equation}
By using the fusion rules of $\phi_{(1,2)}$ and $\phi_{(1,3)}$ , \cite{Antunes:2022vtb} showed that it is consistent under RG to truncate only to deformations by weakly relevant operators of the form
\begin{equation}
    \epsilon_i \equiv \phi_{(1,3)}^{(i)}\,, \qquad \sigma_{ijkl}\equiv \phi_{(1,2)}^{(i)}\phi_{(1,2)}^{(j)}\phi_{(1,2)}^{(k)}\phi_{(1,2)}^{(l)}\,,    
\end{equation}
and proceeded to study the $S_N$ invariant action, with the deforming operators $\epsilon= \sum_i \epsilon_i$ and $\sigma=\sum_{i<j<k<l} \sigma_{ijkl}$. We will consider the most general possible formal action of the form
\begin{align}
    S= \sum_i S_m^{(i)}\!+ \!\int d^2x \sum_ig^i\epsilon_i +  \!\int d^2x \!\!\!\!\sum_{i<j<k<l}\!\!\!\!g^{ijkl} \sigma_{ijkl} \end{align}
    which, when no symmetries are imposed has $N+\binom{N}{4}$ different couplings. The corresponding  one-loop RG equations are given by
\begin{align}
\label{betaeqs}
    \beta_i&= \frac{4}{m}g_i - \frac{4 \pi}{\sqrt{3}}g_i^2-\frac{\sqrt{3}\pi}{2}\sum_{j<k<l} g^2_{ijkl}\,,\nonumber\\
    \beta_{ijkl} &= \frac{6}{m}g_{ijkl} - \sqrt{3}\pi (g_i+g_j+g_k+g_l)g_{ijkl}\nonumber\\
    &-2\pi \sum_{n<m} g_{\{ijnm\}}g_{\{klnm\}}\,,
\end{align}
    where $\{ijkl\}$ denotes the (unique) appropriately ordered permutation of the quadruplet $\{i,j,k,l\}$. Note that the last term is only non-vanishing for $N\geq6$.
    To make these equations manageable, it is natural to impose symmetries $H\subset G=S_N$ acting on the indices $\{i,j,k,l\}$. However, if this is considered in full generality, solving these equations remains a daunting task because of the following 
    
\medskip

    \textbf{Theorem 1} (Cayley): \textit{Every finite group $H$ can be realized as a subgroup of a permutation group $S_N$ for sufficiently large $N$.}
    
\medskip  
The proof is elementary and can be found in any abstract algebra textbook (see e.g. \cite{DummitFoote:Algebra}). 
This result means that the space of solutions is potentially bigger than the space of all finite groups. A useful organizing principle is provided by 

\medskip

    \textbf{Theorem 2} (O'Nan-Scott \cite{Scott:1980reps}): \textit{The maximal subgroups of $S_N$ are either}
    \begin{itemize}
        \item $S_M\times S_{N-M}$
        \item $S_{N/M}\rtimes(S_{M})^{N/M}$
        \item \textit{Primitive subgroups (do not preserve any partition of} $N$) 
    \end{itemize}
 where a maximal subgroup $H$ is defined such that there is no proper subgroup $K\subset G=S_N$ such that $H\subset K$. The so-called primitive subgroups of $S_N$ are further subdivided into four classes, but we find it more convenient to invoke a more general result
 
\medskip 

\textbf{Theorem 3} (Classification of finite simple groups \cite{GLS:CFSG}): \textit{Every finite simple group is isomorphic to one of the following}
 \begin{itemize}
        \item \textit{A cyclic group} $\mathbb{Z}_p$ \textit{of prime order} $p$
        \item \textit{An alternating group} $A_N$ \textit{with} $N\geq5$
        \item \textit{A group of Lie type (16 infinite families)} 
        \item  \textit{One of the 26 sporadic groups}
    \end{itemize}

\medskip 

where we recall that $A_N$ is the even permutation subgroup of $S_N$ and that a group is simple when it has no proper normal subgroups (subgroups invariant under conjugation in $G$). The sporadic finite groups, and in particular the largest of them all, the \textit{Monster} group, famously appear in the context of chiral 2d CFTs with $c=24$ \cite{Lin:2019hks} playing a role in the so-called monstrous moonshine \cite{Conway:1979qga}. While mathematically speaking, the way the symmetry is realized on a lattice of 24 bosons is rather striking, physically speaking this is a symmetry of a \textit{free} model. Instead, we propose that such symmetries should also be taken seriously for \textit{interacting} theories, as they can be realized as symmetries of the one-loop fixed point equations \eqref{betaeqs} (which are admittedly much more prosaic mathematically). Similarly, finite groups of Lie type are often overlooked as symmetries of physical systems in general and CFTs in particular, and we will see that they can also be naturally realized in our setup.

{\bf Classification for $N=4$ and large $m$ conformal manifolds.}
Let us begin with the simplest case $N=4$, which has a single $\sigma$-type coupling and four $\epsilon$-type couplings. The beta function equations read
\begin{equation}
    \begin{aligned}
        &\beta_{i} = \frac{4}{m}g_{i}-\frac{4\pi}{\sqrt{3}}g_{i}^2-\frac{\sqrt{3}\pi}{2}g_{\sigma}^2=0\,,\\
        &\beta_{\sigma}=\frac{6}{m}g_{\sigma}-\sqrt{3}\pi(\sum_{i=1}^4g_{i})g_{\sigma}=0\,,
    \end{aligned}
\end{equation}
where we used the shorthand notation $g_\sigma\equiv g_{1234}$. There are decoupled fixed points at $(g_{i}^*,g_{\sigma}^*)=(\frac{\sqrt{3}}{\pi m},0)$, corresponding to 4 copies of the Zamolodchikov flow.  Solving $\beta_{\sigma}=0$ assuming $g_{\sigma}\neq0$ and replacing $g_{i}\to \frac{\sqrt{3}}{2\pi m}+h_{i}$ yields $\sum_{i=1}^4h_{i}= 0$. 
Furthermore, the first four beta functions can be recast as $h_{i}=\pm\frac{\sqrt{3}}{2\pi m}\sqrt{1-\frac{m^2\pi^2}{2}g_{\sigma}^2}$.
Therefore we have a continuous family of unitary solutions with 
\begin{align}
    g_{i}^*=\frac{\sqrt{3}}{2\pi m}\left(1\pm\sqrt{1-\frac{m^2\pi^2}{2}g_{\sigma}^2}\right)\,,
\end{align} where the plus sign is taken for two of the $g_{i}$, while the other two have the opposite signs, and $0<|g_{\sigma}|\leq\frac{\sqrt{2}}{m\pi}$. Such fixed points have manifest $\mathbb{Z}_2\times \mathbb{Z}_2$ symmetry and in the special cases $g_{\sigma}=\pm\frac{\sqrt{2}}{m\pi}$ then all $g_{i}$ coincide at $g_{i}^*=\frac{\sqrt{3}}{2\pi m}$, yielding the two $S_4$ fixed points discovered in \cite{Antunes:2022vtb}. The $S_4$ symmetric theories were shown to be non-generic, exhibiting two-loop conserved currents as well as certain special non-invertible symmetries. The presence of these large $m$ conformal manifolds is not surprising (see for instance \cite{Benedetti:2019eyl,Antunes:2024mfb} for other examples), but they are expected to lift at subleading order, which we will now procceed to show. 

First we recall that the two-loop beta function in conformal perturbation theory reads (with no sum over $I$)
\begin{equation}
           \beta^I=-\delta_Ig^I-\pi C^{I}_{JK}g^Jg^K+c^I_{JKL}g^Jg^Kg^L\,,
\end{equation}
where $I$ labels all possible couplings, the tree level term is controlled by $\delta_I\equiv2-\Delta_I$, the one-loop term is controlled by the OPE coefficent $C^{I}_{JK}$ and the two-loop term is controlled by a regularized integrated four-point function
\begin{equation}
    c^I_{JKL} =\int_V\ud^2x\ud^2y\langle O_J(0)O_K(x)O_L(y)O_I(\infty)\rangle_{\textbf{reg}}\,,
\end{equation}
and we discuss the regularization scheme in  Appendix A. Integrating all the necessary products of four point-functions of $\phi_{(1,3)}$ and $\phi_{(1,2)}$, which simplify in the large $m$ limit, we arrive at the two-loop beta-function equations
\small
    \begin{align}
    \label{twoloop}
        &\beta_{\sigma}=\delta_{\sigma}g_{\sigma}-\sqrt{3}\pi \left(\sum_{i=1}^4g_{i}\right)g_{\sigma}-\frac{9}{2}\pi^2g_{\sigma}^3-6\pi^2g_{\sigma}\left(\sum_{i=1}^4g_{i}^2\right)\,,\nonumber\\
        &\beta_{i}=\delta_{i}g_{i}-\frac{4\pi}{\sqrt{3}}g_{i}^2-\frac{\sqrt{3}\pi}{2}g_{\sigma}^2-6\pi^2g_{\sigma}^2g_{i}-16\pi^2g_{i}^3\,.
    \end{align}
    \normalsize
 As expected, the one-loop manifolds of $\mathbb{Z}_2\times\mathbb{Z}_2$ fixed points are lifted and in fact, the only remaining solutions are the $S_4$ fixed points with a corrected value of the critical coupling.
 
{\bf Classification for $N=5$.} 
For five copies, there are 5 couplings of each type, making the equations much more non-trivial. Similarly to the $N=4$ case we can perform the shift $g_{i}\to \frac{\sqrt{3}}{2\pi m}+h_{i}$ to obtain:
\begin{equation}
    \begin{aligned}
        &\beta_{g_{i}} =\frac{\sqrt{3}}{\pi} -\frac{4\pi}{\sqrt{3}}h_{i}^2-\frac{\sqrt{3}\pi}{2}\sum_{j<k<l}g_{ijkl}^2\,,\\
        &\beta_{ijkl}=-\sqrt{3}\pi(h_{i}+h_{j}+h_{k}+h_{l})g_{ijkl}\,.
    \end{aligned}
\end{equation}
The fully decoupled fixed points (products of $\mathcal{M}_m$ and $\mathcal{M}_{m-1}$) are those with all $g_{ijkl}=0$. Each corresponds to a choice of turning on or not each of the 5 $g_{i}$, resulting in $6$ decoupled fixed points. Turning on only one of the $g_{ijkl}$ yields all the non-decoupled fixed points of $N=4$. For each choice of $g_{ijkl}$ there are 2 fixed points, times 2 for the turning on or off of the extra $\epsilon^i$, resulting in 4 distinct fixed points which are products of an interacting $N=4$ theory with a single minimal model for each $g_{\sigma_{ijkl}}$.

Things become more interesting when 2 $g_{ijkl}$ are turned on. In this case, there are 8 solutions with $S_3\times \mathbb{Z}_2$ symmetry up to permutation , i.e.
\begin{equation}
    \begin{aligned}
        &g_{1345}^*\,,\,g_{2345}^*=\pm \frac{4}{\sqrt{17\pi}m}\,,\quad g_{1234}^*=g_{1235}^*= g_{1245}^*=0\,,\\
        &h_{1}=h_{2}=-3h_{3}=-3h_{4}=-3h_{5}=\pm\frac{3}{2\pi m}\sqrt{\frac{3}{17}}\,,
    \end{aligned}
\end{equation}
and the symmetry pattern ensures these are genuinely new fixed points. Turning on 3 or 4 $g_{\sigma^{ijkl}}$ does not yields unitary fixed points. Finally, when all $g_{ijkl}$ are turned on, all $h^*_{i}=0$ and $g^*_{ijkl}=\pm\frac{1}{\sqrt{2}\pi m}$, yielding $6$ fixed points. Depending on the signs, the symmetry group is either the full $S_5$, a non-decoupled $S_4$ or $S_3\times \mathbb{Z}_2$.
To summarize (indicating $h_{i}$ instead of $g_{i}$):

\begin{table}[h]
\footnotesize
    \centering
    \begin{tabular}{|c|c|c|c|}
    \hline
       $(h_i^*,g_{ijkl}^*)$ & \# & $H$ \\
    \hline
      $\pm\frac{\sqrt{3}}{2\pi m}(1,1,1,1,1,0,0,0,0,0)$& 2 & $S_5$\\
    \hline
      $\pm\frac{\sqrt{3}}{2\pi m}(-1,1,1,1,1,0,0,0,0,0)$& 2 & $S_4$ \\
    \hline
      $\pm\frac{\sqrt{3}}{2\pi m}(-1,-1,1,1,1,0,0,0,0,0)$& 2 & $S_3\times \mathbb{Z}_2$ \\
    \hline
    $(h_+,h_+,h_-,h_-,\frac{\sqrt{3}}{4\pi m}(1\pm1),g_{\sigma},0,0,0,0)$& 4\footnote{for each $g_{\sigma}$} & $\mathbb{Z}_2\times \mathbb{Z}_2$ \\
    \hline
    $\pm\frac{\sqrt{3}}{6\sqrt{17}\pi m}(9,9,-3,-3,-3,0,0,0,\pm 8\sqrt{3},\pm 8\sqrt{3})$& 8 & $S_3\times \mathbb{Z}_2$ \\
    \hline
    $\pm\frac{1}{\sqrt{2}\pi m}(0,0,0,0,0,1,1,1,1,1)$& 2 & $S_5$ \\
    \hline
    $\pm\frac{1}{\sqrt{2}\pi m}(0,0,0,0,0,-1,1,1,1,1)$& 2 & $S_4$ \\
    \hline
    $\pm\frac{1}{\sqrt{2}\pi m}(0,0,0,0,0,1,1,-1,-1,-1)$& 2 & $S_3\times \mathbb{Z}_2$ \\
    \hline
    \end{tabular}
\end{table}

{\bf Explorations for $N\geq6$}
Rigorously classifying all fixed points for $N\geq6$ becomes out of reach, and we will be content to go through subgroups $H\subset S_6$ that are not too small, i.e. do not have too many independent couplings. Apart from the $S_6$ symmetric solutions we were able to classify fixed points with the symmetries in Table I below.
\begin{table}[h]
    \centering
    \footnotesize
    \begin{tabular}{|c|c|c|}
    \hline
       $H$ & $\#$ of coupl.& $\#$ of fixed pts.\\
    \hline
     $S_5$ & 4 & 10\\
    \hline
    $(S_3\times S_3)\rtimes \mathbb{Z}_2$ & 3 & 4\\
    \hline
    $S_4 \times \mathbb{Z}_2$ & 5& 22 \\
    \hline
    $S_3 \times S_3$ & 5& 16 \\
    \hline
    $S_4 $ & 7 & 26\\
    \hline
    $D_{10} $ & 5 & 2\\
    \hline
    \end{tabular}
    \label{tab:n6syms}
    \caption{Fixed points for $N=6$ coupled minimal models.}
\end{table}

To generate the beta-function equations for these subgroups, we select a set of generators of $H$ and find an orbit of couplings closed under the action of $H$. Each orbit leads to an independent coupling. 
In particular, we find exactly subgroups of the form predicted by \textbf{Theorem 2}, preserving certain integer partitions of $6$. We find particularly remarkable the existence of fixed points with $S_5$ symmetry realized with 6 theories in a non-trivial way. Similar fixed points were also detected for Wilson-Fisher type models \cite{Osborn:2020cnf} (realizing $O(M)$ symmetry with $N>M$ scalars), but have not been exhaustively explored. We also note the existence of fixed points with $D_{10}$ symmetry: this is the dihedral group of a regular pentagon, containing a $\mathbb{Z}_5$ cyclic rotation and $\mathbb{Z}_2$ reflections. This cyclic symmetry is particularly appealing in the large $N$ limit: it is precisely the symmetry expected for an emergent $3$rd dimension. An exhaustive presentation of the set of $H$ invariant couplings and the associated fixed points is given in Appendix B.

Partition preserving subgroups for $N\geq7$ can systematically be found in a similar way. However, starting at this value of $N$ we also encounter finite groups which are not of this form. In particular, the first finite simple group of Lie type occurs since
\begin{align}
   & \textrm{PSL}_2(7) = \\
&\left\{
\begin{pmatrix}
a & b \\
c & d
\end{pmatrix}
\,\middle|\,
a,b,c,d \in \mathbb{Z}/7\mathbb{Z},\;
ad - bc = 1
\right\}/\left\{I,-I\right\} \nonumber
\end{align}
is a subgroup of $S_7$. This is the projective linear group valued in the field of integers $mod$ 7 and is a Chevalley group (a Lie group over a finite field) with 168 elements \cite{Note1}. To find fixed points invariant under this group we use the following algorithm which can be adapted to any subgroup $H$ (we utilize the GAP library \cite{SmallGrp:GAP} to implement this procedure):
\begin{itemize}
    \item Identify the (in this case two) generators of $H=\textrm{PSL}_2(7)$ as a set of permutations in $S_7$.
    \item Generate the full group by composition of the generators.
    \item Start from a coupling and find its orbit.
    \item Repeat until all couplings have been exhausted.
    \item Derive the necessary combinatorial factors to fix the coefficients in the beta function equations.
\end{itemize}
Using this procedure we find that there are actually only three independent couplings, making the equations completely tractable. There are a total of 6 fixed points, including the 4 known $S_7$ fixed points. The 2 new fixed points are
\begin{equation}
    g_{\epsilon} = \frac{\sqrt{3}\pm 1}{2\pi m}\,,\qquad g_{\sigma^1} = \mp \frac{1}{\sqrt{3}\pi m}\,,\quad g_{\sigma^2}=0\,,
\end{equation}
where the couplings are defined in the Appendix C.
$\rm{PSL}_2(7)$ is maximal in $A_7$, therefore these 2 new fixed points are not invariant under a larger group. We also note that $\rm{PSL}_2(7)$ is isomorphic to $\rm{PSL}_3(2)$.

We can proceed to find a few more fixed points with larger $N$ and symmetry of Lie type. For $N=11$ we were able to find and classify $\rm{PSL}_2(11)$ fixed points. 
The $\sigma^{ijkl}$ couplings fall into three singlets, while the $\epsilon^i$ form only one yielding 4 couplings in total. There are a total of 16 fixed points, but only 8 are real, including the 4 known $S_{11}$ fixed points. 2 of the new real fixed points are
\begin{align}
    g_{\epsilon} &= \frac{\sqrt{3}}{2\pi m}\pm \frac{9} {2\sqrt{47}\pi m}\,, \nonumber\\
    \, g_{\sigma^1} &=-g_{\sigma^2} =-g_{\sigma^3} = \pm\frac{ 1}{\sqrt{141}\pi m}\,.
\end{align}
The other two contain roots of 10th order polynomials, but numerically approximate to
$g_{\epsilon} = \frac{\sqrt{3}}{2\pi m}\pm \frac{0.661}{\pi m}\,,\, g_{\sigma^1}=\pm \frac{0.110}{\pi m}\,,\,  g_{\sigma^2} = \mp \frac{0.069}{\pi m}\,,\, g_{\sigma^3} = \mp \frac{0.056}{\pi m}$.
Since $\rm{PSL}_2(11)$ is maximal in $M_{11}$ (a Mathieu group which will be discussed below) and $M_{11}$ is maximal in $A_{11}$, these 4 new fixed points are not invariant under a larger group. For $PSL_2(13)$ the $\sigma^{ijkl}$ fall into four singlets, while the $\epsilon^i$ into one. There are consequently 5 couplings. There are a total of 32 fixed points, but only 20 are real, including the 4 known $S_{14}$ fixed points. The remaining fixed points are non-trivial but can only be obtained numerically. Since $PSL_2(13)$ is maximal in $A_{14}$, these new fixed points are not invariant under a larger group. 
 
 \medskip

The remaining finite groups of Lie type are either:  large enough that their generators don't lie in the GAP small group library; non-simple groups; or groups isomorphic to simple groups that are not of Lie type (e.g. $A_N$). Therefore, a more systematic exploration of groups of Lie type (which includes the aforementioned Chevalley groups \cite{Chevalley:1955} as well as Steinberg groups \cite{Steinberg:1961} and Suzuki-Ree groups \cite{Suzuki:1960,Ree:1960,Ree:1961}) requires more powerful group theory tools and is left for future work. 

\medskip

 As \textbf{Theorem 3} shows, finite simple groups do not end in the 16 Lie type families. In fact, the most exotic groups $H\subset S_N$ fall in the fourth category of \textbf{Theorem 3}: they are one of 26 sporadic groups. The simplest ones are the so-called Mathieu Groups \cite{Mathieu:1861,Conway:Atlas}. All 5 of them have known generators as subgroups of $S_N$ implemented e.g. in \texttt{Mathematica}. One can then find if the $\sigma^{ijkl}$ couplings transform in more than one singlet for each of those groups. As it turns out, it is only possible to construct the $S_N$ singlet for $M_{11}$, $ M_{12}$ and $M_{23}$ with $N=11,12,23$ respectively. For $M_{22}$ there are actually 2 singlets, and correspondingly 4 new fixed points of the $\beta$ function equations  
\begin{equation}
    \begin{aligned}
        &h_{\eps}^2+\frac{3}{4}(560g_{\sigma^1}^2+105g_{\sigma^2}^2)=\frac{3}{4\pi^2m^2}\,,\\
        &2\sqrt{3}g_{\sigma^1}h_{\eps}+324g_{\sigma^1}^2+9g_{\sigma^2}^2+126 g_{\sigma^1}g_{\sigma^2}=0\,,\\
        &2\sqrt{3}g_{\sigma^2}h_{\eps}+336g_{\sigma^1}^2+27g_{\sigma^2}^2+96 g_{\sigma^1}g_{\sigma^2}=0\,,
    \end{aligned}
\end{equation}
which are however all complex. $M_{24}$, having an order of 244 823 040, is too big to be checked in the brute force manner we have outlined above. 
While these Mathieu fixed points are non-unitary, we find it rather remarkable that non-trivial fixed points with this symmetry exist at all. Indeed, in the case $N=22$ where they do exist, the fixed point equations are rather tractable as the number of invariant couplings is quite small. This suggests that with a more refined group theory implementation (which does not simply span the full set of group elements by brute force) it might be possible to find fixed points for $M_{24}$ and even larger sporadic groups.

{\bf Some rigorous results for general $N$.}
Before concluding, we present some sharp results that apply for arbitrary $N$. We begin with the statement that

\medskip
\textit{Fixed points with $A_N$ symmetry always enhance to $S_N$.}
\medskip

The proof is straightforward: Consider a string of 4 distincts integers $s=\{i,j,k,l\}$. The question we need to ask is whether or not for any $s$, there exists $g \in A_N$ such that $g(s_0=\{1,2,3,4\})=s$. Let us assume that we found $s$, such that there exists $g\in S_N\backslash A_N$ with $g(s_0)=s$. Then acting with the group element $g_0=(ij)$ on $s$ would not modify it: $g_0(s)=s$. Consequently, $(g_0 \cdot g)(s_0)=s$, with $g_0\cdot g \in A_N$. We have found a group element in $A_N$ sending $s_0$ to $s$, and similarly this can be done for any $s$ (and $s_0$). Therefore, there can only be one singlet under which the $\sigma^{ijkl}$ transform. A similar argument applies to the $\epsilon^i$, ensuring that the only action we can write down that is invariant under $A_N$ is also invariant under $S_N$.

We then consider fixed points with $H=S_M\times S_{N-M}$ symmetry. Using the methods described above, we find that there are 7 couplings. We present the equations in Appendix D, but we are not able to solve them in full generality. However, specializing to the case $M=N/2$ with $N$ even, we are able to impose an additional $\mathbb{Z}_2$ symetry swapping the two sides of the bipartition. This reduces the number of independent couplings to 4, and we were then able to prove that 

\medskip
\textit{16 real fixed points with $H=(S_{N/2}\times S_{N/2})\rtimes \mathbb{Z}_2$ exist iff $N\leq10$. Otherwise 12 real fixed points exist.}
\medskip

which can be done by examining the explicit (but very cumbersome) solutions which we present in Appendix D. We note that among these solutions there are always fixed points which do not have enhanced symmetry and are not simply a tensor product, i.e. \textit{they are genuinely new fixed points}.

Similarly, we can also consider the subgroup $H=(\mathbb{Z}_{2})^{N/2}\rtimes S_{N/2}$, which acts naturally on $N/2$ pairs of integers of the form $\{i,i+1\}$ \cite{Connor}. In Appendix E we construct the 4 invariant couplings and show that

\medskip
\textit{10 real fixed points with $H=(\mathbb{Z}_{2})^{N/2}\rtimes S_{N/2}$ exist for $N\geq10$.}
\medskip

and also obtain closed form solutions for the fixed points at generic $N$.

Finally, we present an elementary bound on the space of solutions. Such bounds have been derived for Wilson-Fisher fixed points \cite{Rychkov:2018vya,Hogervorst:2020gtc} making use of the perturbative $a-$function of which the beta functions can be obtained as a gradient \cite{Jack:1990eb}. In our case,  we can simply perform the change of variable $g_{i} = \frac{\sqrt{3}}{2\pi m }+h_{i}$, then the $\beta$ function for $g_{i}$ evaluated at a fixed point yields the quadratic constraint
\begin{equation}
    \sum_{i<j<k<l}(g_{ijkl}^*)^2 \leq \frac{N}{2\pi^2m^2}\,,
\end{equation}
which is saturated by solutions with $h_i^*=0$.

{\bf Discussion.}
The space of compact irrational CFTs with only Virasoro symmetry is only superficially understood, if at all. We might have scratched its surface at a few new places, but were far from making a dent. A systematic examination in the spirit of \cite{Rong:2023xhz} remains to be undertaken, and each of the uncovered fixed points deserves a systematic perturbative study of its CFT data to understand patterns in the breaking of its enhanced chiral symmetry \cite{Antunes:2022vtb}, in the general organization of the spectrum into Regge trajectories \cite{Kusuki:2018wpa,Collier:2018exn} and on the assymptotic behavior of OPE coefficients \cite{Collier:2019weq}.

The large $m$ expansion has proven to be a reliable tool in the search for `generic' 2d CFTs, but the large $N$ expansion has so far remained untapped. While the standard Hubbard-Stratanovich trick only works for $O(N)$ symmetric quartic interactions, an $S_N$ symmetric version was shown to work in \cite{Binder:2021vep} in the context of Hypercubic CFTs in 3d, studied in conformal perturbation theory from a two-copy deformation of tensored Ising models. We expect the methods there to be adaptable to our four-copy interaction as well as the more traditional two-copy setups of \cite{djlp98}.

We end by emphasizing that many of the structural properties of non-perturbative CFTs are not a priori verifiable in perturbation theory. While \cite{djlp98,Antunes:2025huk} have provided hints for the non-perturbative existence of `generic' 2d CFTs, the conformal bootstrap remains as the premier tool in the non-perturbative study of CFTs. Some first steps were taken in \cite{Kousvos:2024dlz}, but we expect the modular bootstrap incorporating non-invertible symmetries \cite{Lin:2023uvm,WIP1} to play a key role in placing these CFTs at the same footing as the 3d Ising model: unsolvable but well-understood.

\vspace{10pt}

{\bf Acknowledgements}
We thank Connor Behan, Junchen Rong and Andy Stergiou for useful discussions. AA further thanks Andreia Gon\c{c}alves for continued inspiration. AA and NS are funded by the European Union (ERC, FUNBOOTS, project number 101043588, PI Miguel Paulos). Views and opinions expressed are however those of the authors only and do not necessarily reflect those of the European Union or the European Research Council Executive Agency. Neither the European Union nor the granting authority can be held responsible for them.

\newpage
\onecolumngrid
\appendix

\begin{center}
\textbf{Appendix A: regularization scheme}    
\end{center}

The goal of this appendix is to summarize how to obtain the logarithmically divergent part of integrals of the type
\begin{equation}
\label{eq:div}
    c^I_{JKL} =\int_V\ud^2x\ud^2y\langle O_J(0)O_K(x)O_L(y)O_I(\infty)\rangle_{\textbf{reg}}\,,
\end{equation}
when $m\to\infty$, that is, when operators approaching  marginality are exchanged in the OPE. Such integrals have already been explored in several previous works \cite{Behan:2025ydd,Behan:2017emf,Komargodski:2016auf}. One may worry that the $m\to\infty$ limit is radically different from the $m=\infty$ theory, i.e. the limit and the integral don't commute. To make sure that no such problem arises, we independently computed those divergences numerically and found that \textit{for the theories we consider} (see \cite{Behan:2025ydd} for a case where this is not true), we are allowed to commute integration and limit.\\
We integrate over a finite volume $V$, which serves as an IR regulator, such that we don't capture contributions coming from operators approaching $O_I$. By rescaling $x\to x/|y|$ and $y\to y/|y|$, we can explicitly extract the $\log$ divergence:
\begin{align}
\label{eq:inttemp}
   \int_V\ud^2x\ud^2y\langle O_J(0)O_K(x)O_L(y)O_I(\infty)\rangle = 2\pi \log\left(V/a\right)\int_V \ud^2 x\,\langle O_J(0)O_K(x)O_L(1)O_I(\infty)\rangle\,.
\end{align}

The right hand side of (\ref{eq:inttemp}) still contains divergences coming from the exchange of relevant and marginal operator between $O_J$, $O_K$ and $O_L$. We should remove them explicitly, then integrate:  
\begin{equation}
    c^I_{JKL} =2\pi\int_V\ud^2x\langle O_J(0)O_K(x)O_L(1)O_I(\infty)\rangle_{\textbf{reg}}\,.
\end{equation}

Let us exemplify this by doing the computation for the $N=4$ case. In the limit $m\to\infty$, the correlators take a simple form:
\begin{equation}
    \begin{aligned}
        &G_{\sigma\sigma\sigma\sigma}(x) = \left(\frac{1-\Re e(x)+|x|^2}{|x||1-x|}\right)^4\,,\\
        &G_{\sigma\sigma\eps_i\eps_i}(x) = \frac{4-2\Re e(x)+8\Re e(x)^2-3|x|\Re e(x)+8|x|^2+3|x|^4}{4|x|^4|1-x|^2}\,,\\
        &G_{\eps_i\eps_i\eps_i\eps_i}(x) = \frac{3-12 \Re e(x)+36\Re e(x)^2-48\Re e(x)^3+16\Re e(x)^4}{3|x|^4|1-x|^4}\\
        &+ \frac{10|x|^2-48\Re e(x)|x|^2+96\Re e(x)^2|x|^2-48\Re e(x)^3|x|^2}{3|x|^4|1-x|^4}\\
        &+ \frac{9|x|^4-48\Re e(x) |x|^4+36\Re e(x)^2|x|^4+10|x|^6-12\Re e(x)|x|^6+3|x|^8}{3|x|^4|1-x|^4}\,,
    \end{aligned}
\end{equation}
and the $i$ indices should not be summed. Correlations functions involving different $\eps_i$ are pure powers, thus producing no finite contribution to $c^I_{JKL}$. The regulated correlators are obtained by removing the divergences from the $x\to 0$, $x\to 1$ and $x\to\infty$ limits:

\begin{equation}
    \begin{aligned}
        &\Tilde{G}_{\sigma\sigma\sigma\sigma}(x)=G_{\sigma\sigma\sigma\sigma}(x)-\left(\frac{1}{|x|^4}+1+\frac{1}{|1-x|^4}+\frac{3}{2}\left(\frac{1}{|x|^2}+\frac{1}{|x|^2|1-x|^2}+\frac{1}{|1-x|^2}\right)\right)\,,\\
        &\Tilde G_{\sigma\sigma\eps_i\eps_i}(x)=G_{\sigma\sigma\eps_i\eps_i}(x) -\left(\frac{1}{|x|^4}+\frac{1}{|x|^2}+\frac{1}{|x|^2|1-x|^2}-\frac{1}{4|1-x|^2}\right)\,,\\
        &\Tilde G_{\eps_i\eps_i\eps_i\eps_i}(x)=G_{\eps_i\eps_i\eps_i\eps_i}(x)-\left(\frac{1}{|x|^4}+1+\frac{1}{|x-1|^4}+\frac{16}{6}\left(\frac{1}{|x|^2}+\frac{1}{|x|^2|1-x|^2}+\frac{1}{|1-x|^2}\right)\right)\,.
    \end{aligned}
\end{equation}
One can check that the coefficients in front of each power match the OPE coefficients of the exchanged operators. Each integral can be done explicitly. In practice, we go to radial coordinates and integrate separately the regions $r>1$ and $r<1$. If we are not careful, we may miss contributions from the $x,\bar{x}\to 1$ region. As a rule of thumb, each contributions of $1/(x-1)^2+1/(\bar{x}-1)^2$ near $x,\bar{x}\to 1$ (the only relevant divergences in our case) will contribute a total of $-2\pi$. See Appendix D of [63] for more details. Once the dust settles, we find:
\begin{equation}
    \begin{aligned}
        &\int_V\ud ^2x \, \Tilde{G}_{\sigma\sigma\sigma\sigma}(x)=-\frac{9\pi}{4}\,,\\
        &\int_V\ud ^2x\,  \Tilde{G}_{\sigma\sigma\eps_i\eps_i}(x)=-\pi\,,\\
        &\int_V\ud ^2x \, \Tilde{G}_{\eps_i\eps_i\eps_i\eps_i}(x)=-8\pi\,.
    \end{aligned}
\end{equation}

and the final result is
\begin{equation}
    \begin{aligned}
        c^{\sigma}_{\sigma\sigma\sigma}=-\frac{9\pi^2}{2}\,,\qquad c^{\sigma}_{\epsilon^i\epsilon^i\sigma}=c^{\epsilon^i}_{\epsilon^i\sigma\sigma}=-2\pi^2\,,\quad c^{\epsilon^i}_{\epsilon^i\epsilon^i\epsilon^i}=-16\pi^2\,,
    \end{aligned}
\end{equation}
from which we obtain the two-loop beta-function equations \eqref{twoloop}.
\begin{center}
\textbf{Appendix B: invariant couplings and fixed points for $N=6$}    
\end{center}

We summarize the list of invariant couplings and fixed points for all subgroups $H\subset G$ that we studied. That includes all subgroups with $|H|\geq10$ with the exception of $D_{12}$ \cite{NoteD12}.\\
For any $H$, the 4 $S_6$ fixed points will be part of the set of solutions. We remind the reader that they are given by:
\begin{equation}
    (h_{i},g_{ijkl})\in\left\{\left(\frac{1}{2\sqrt{2}\pi m}\,,-\frac{1}{\sqrt{6}\pi m}\right),\left(-\frac{1}{2\sqrt{2}\pi m}\,,\frac{1}{\sqrt{6}\pi m}\right),\left(\frac{\sqrt{3}}{2\pi m}\,,0\right),\left(-\frac{\sqrt{3}}{2\pi m}\,,0\right)\right\}\,,
\end{equation}
the latter two being the $\mathcal{M}_m^6\to\mathcal{M}_{m-1}^6$ flow and the trivial solution $(g_i,g_{ijkl})=(0,0)$ respectively.
More generally, any fixed point of $H^+$ with $H\subseteq H^+\subseteq S_6$ will be a fixed point of $H$.
We will omit fixed points of $H^+\neq H$ onwards for the sake of readability.\\
Fully decoupled fixed points are solutions for which all $g_{ijkl}=0$. In that case each $h_i$ can take the value $\pm\frac{\sqrt{3}}{2\pi m}$ with independent signs. The number of $h_i$ with a given sign determines the symmetry of such solutions which are always of the form $S_{N-M}\times S_M$, which for $N=6$ yields $S_3\times S_3$, $S_4\times \mathbb{Z}_2$, $S_5$ or $S_6$. However, such solutions always appear as long as $H$ is a subgroup of these groups. We will similarly omit them henceforth for the sake of readability.\\
Finally, below we will only consider fixed point given a presentation of $H$, corresponding to a set of its generators. Equivalent fixed points may be found by acting with $S_N$ elements, or equivalently changing the choice of generators. We will not count these, as no new fixed point is generated this way. Sometimes, it may happen that even for a given choice of generators, a residual symmetry remains, which may result in overcounting fixed points (e.g. $S_3\times S_3$, which contains a $\mathbb{Z}_2$ symmetry that exchange the two $S_3$). We will similarly only count fixed points up to such symmetries.\\
Many fixed points come in sets differing from each other by changes of signs. To make clear which signs are independent from each other, identically colored (black, red or blue) $\pm$'s are not independent. \\

$\bullet$ \underline{$H=S_5$}\\

$S_5\subset S_6$ can be realized as the permutation of indices $i=1,\dots,5$, leaving $i=6$ invariant. Therefore we have 4 invariant operators:
\begin{equation}
    \begin{aligned}
        \epsilon\equiv\sum_{i=1}^5\epsilon_i\,,\quad\epsilon_6\,,\quad\sigma_1\equiv\sum_{1\leq i,j,k,l\leq5}\sigma_{ijkl}\,,\quad \sigma_2 \equiv\sum_{1\leq i,j,k\leq 5}\sigma_{ijk6}\,.
    \end{aligned}
\end{equation}
There are 16 real fixed points, including 10 interacting fixed points with $H^+=S_5$. 2 of them are almost identical to the interacting $S_6$ solutions, but with the sign of $g_{\sigma_2}$ flipped. 4 are the tensor product of an interacting $N=5$ fixed point with one copy of the flow $\mathcal{M}_m\to\mathcal{M}_{m-1}$/trivial solution.

\begin{table}[h]
    \centering
    \footnotesize
    \begin{tabular}{|c|c|c|}
    \hline
       $\left(h_{\epsilon}^*,h_{\epsilon_6}^*,g_{\sigma_1}^*,g_{\sigma_2}^*\right)$ & Degeneracy & Tensor product\\
    \hline
     $\pm\left(\frac{7}{2\sqrt{55}\pi m},-\frac{1}{2\pi m}\sqrt{\frac{5}{11}},-\frac{4}{\sqrt{165}\pi m},\textcolor{red}{\pm}\frac{2}{\pi m}\sqrt{\frac{7}{165}}\right)$&4&No\\
    \hline
     $\pm\left(\frac{1}{2\sqrt{2}\pi m},\frac{1}{2\sqrt{2}\pi m},-\frac{1}{\sqrt{6}\pi m},\frac{1}{\sqrt{6}\pi m}\right)$&2&No\\
     \hline
      $\pm\left(0,\frac{\sqrt{3}}{2\pi m},\textcolor{red}{\pm}\frac{1}{\sqrt{2}\pi m},0\right)$&4&Yes\\
     \hline
    \end{tabular}
    \caption{$H=S_5$: interacting fixed points}
\end{table}

$\bullet$ \underline{$H=(S_3\times S_3)\rtimes \mathbb{Z}_2$}\\

This group of 72 elements can be realized as the group preserving $\{1,2,3\}$ and $\{4,5,6\}$ along with the exchange $\{1,2,3\}\leftrightarrow \{4,5,6\}$. There are only 3 invariant operators:
\begin{equation}
\label{eq:s3s3z2}
    \begin{aligned}
        \epsilon\equiv\sum_{i=1}^6\eps_i\,,\quad \sigma_1\equiv\sum_{i=1}^3\sigma_{i456}+\sum_{i=4}^6\sigma_{123i}\,,\quad \sigma_2\equiv\sum_{i,j=1}^3\sum_{k,l=4}^6\sigma_{ijkl}\,.
    \end{aligned}
\end{equation}
These can be seen as a special case of the more general $(S_{N/2}\times S_{N/2})\rtimes \mathbb{Z}_2$, which has one less coupling than the generic case, due to the fact that one cannot build operators invariant under the action of only one of the $S_3$.\\

There are 7 solutions, including 4 interacting fixed points with $H^+=(S_3\times S_3)\rtimes \mathbb{Z}_2$. Similarly to $H=S_5$, we find that 2 of those fixed points are almost identical to the interacting $S_6$ fixed points, differing only in the sign of $g_{\sigma_1}$.\\

\begin{table}[h]
    \centering
    \footnotesize
    \begin{tabular}{|c|c|c|}
    \hline
       $\left(h_{\epsilon}^*,g_{\sigma_1}^*,g_{\sigma_2}^*\right)$ & Degeneracy & Tensor product\\
    \hline
     $\pm\left(\frac{\sqrt{3}}{\sqrt{31}\pi m},0,-\frac{3}{\sqrt{31}\pi m}\right)$&2&No\\
    \hline
     $\pm\left(\frac{1}{2\sqrt{2}\pi m},\frac{1}{\sqrt{6}\pi m},-\frac{1}{\sqrt{6}\pi m}\right)$&2&No\\
     \hline
    \end{tabular}
    \caption{$H=(S_3\times S_3)\rtimes \mathbb{Z}_2$: interacting fixed points}
\end{table}

$\bullet$ \underline{$H=S_4\times \mathbb{Z}_2$}\\

$S_4\times \mathbb{Z}_2$ can be realized as the set of permutations leaving $\{1,2,3,4\}$ and $\{5,6\}$ invariant. There are consequently 5 invariant operators:
\begin{equation}
    \begin{aligned}
        \eps^1\equiv\sum_{i=1,4}\eps_i\,,\quad \eps^2\equiv\sum_{i=5,6}\eps_i\,,\quad \sigma_1 \equiv \sigma_{1234}\,,\quad \sigma_2\equiv\sum_{1\leq i,j,k \leq 4}\sum_{l=5}^6\sigma_{ijkl}\,,\quad \sigma_3\equiv\sum_{1\leq i,j\leq 4}\sum_{5\leq k,l\leq 6}\sigma_{ijkl}\,.
    \end{aligned}
\end{equation}

There are 28 fixed points, including 22 interacting fixed points with $H^+=S_4\times \mathbb{Z}_2$. We once agin find again 2 fixed points related to the $S_6$ interacting solutions by a sign flip of $g_{\sigma_2}$. Similarly to the case $H=S_5$ we find 4 solutions that are the tensor product of an interacting $N=4$ fixed point with two copies of the flow $\mathcal{M}_m\to\mathcal{M}_{m-1}$/trivial solution.

\begin{table}[h]
    \centering
    \footnotesize
    \begin{tabular}{|c|c|c|}
    \hline
       $\left(h_{\epsilon^1}^*,h_{\epsilon^2}^*,g_{\sigma_1}^*,g_{\sigma_2}^*,g_{\sigma_3}^*\right)$ & Degeneracy & Tensor Product\\
     \hline
     $\pm\left(\frac{1}{2\pi m}\sqrt{\frac{3}{10}},\frac{1}{2\pi m}\sqrt{\frac{3}{10}},\textcolor{red}{\pm}\frac{3}{\sqrt{10}\pi m},0,\frac{1}{\pi m}\sqrt{\frac{3}{10}}\right)$&4&No\\
     \hline
     $\pm\left(h_{\epsilon^1}(\alpha_i),h_{\epsilon^2}(\alpha_i),g_{\sigma_1}(\alpha_i),\textcolor{red}{\pm}g_{\sigma_2}(\alpha_i),g_{\sigma_3}(\alpha_i)\right)$&12&No\\
     \hline
     $\pm\left(\frac{1}{2\sqrt{2}\pi m},\frac{1}{2\sqrt{2}\pi m},-\frac{1}{\sqrt{6}\pi m},\frac{1}{\sqrt{6}\pi m},-\frac{1}{\sqrt{6}\pi m}\right)$&2&No\\
     \hline
     $\pm\left(0,\frac{\sqrt{3}}{2\pi m},\textcolor{red}{\pm}\frac{\sqrt{2}}{\pi m},0,0\right)$&4&Yes\\
      \hline
    \end{tabular}
    \caption{$H=S_4\times \mathbb{Z}_2$: fixed points}
\end{table}
\small
where
\begin{equation}
    \begin{aligned}
       & h_{\epsilon^1}(x)=\frac{\sqrt{3}}{1018422064\pi m}(-698593081x^{1/2} +45560796023808 x^{3/2} - 
  3163417763904x^{5/2} +93918832768x^{7/2})\,,\\
       & h_{\epsilon^2}(x)=\frac{\sqrt{3}}{1018422064\pi m}(6169467499x^{1/2} -281756498304 x^{3/2} +9490253291712x^{5/2} -136682388071424x^{7/2})\,,\\
       &g_{\sigma_1}(x)=\frac{1}{63651379\pi m}(2818564576x^{1/2}-380510705327x^{3/2}+12319973218800x^{5/2}-117228861008064x^{7/2})\,,\\
       &g_{\sigma_2}(x)=\frac{1}{\sqrt{127302758}\pi m}\sqrt{65960944-8571696277x+349842255984x^2-4077455623488x^3}\,,\\
       &g_{\sigma_3}(x)=-\frac{2}{\pi m}x\,,
    \end{aligned}
\end{equation}
and $\alpha_i$ is one of the three roots of $P_3(x)= 10860096x^3-804400x^2+14241x-64$.\\
\normalsize\\

$\bullet$ \underline{$H=S_3\times S_3$}\\

$H=S_3\times S_3$ is very similar to $H=(S_3\times S_3)\rtimes \mathbb{Z}_2$, being realised as the permutations leaving $\{1,2,3\}$ and $\{4,5,6\}$ invariant. The operators $\eps$ and $\sigma_1$ defined in (\ref{eq:s3s3z2}) are split into two:
\begin{equation}
\label{eq:s3s3}
    \begin{aligned}
        \eps^1=\sum_{i=1}^3\eps_i\,,\quad  \eps^2=\sum_{i=4}^6\eps_i\,,\quad \sigma_{1,1} = \sum_{i=1}^3\sigma_{i456}\,,\quad \sigma_{1,2} = \sum_{i=4}^6\sigma_{123i}\,.
    \end{aligned}
\end{equation}

There are 25 fixed points, including 16 interacting fixed points with $H^+ = S_3\times S_3$. \\

\begin{table}[h]
    \centering
    \footnotesize
    \begin{tabular}{|c|c|c|}
    \hline
       $\left(h_{\eps^1}^*,h_{\eps^2}^*,g_{\sigma_{1,1}}^*,g_{\sigma_{1,2}}^*,g_{\sigma_2}^*\right)$ & \# & Tensor Product\\
     \hline
     $\pm\left(\frac{\sqrt{3}}{2\sqrt{14}\pi m},\frac{\sqrt{3}}{2\sqrt{14}\pi m},\textcolor{red}{\pm}\frac{\sqrt{5}}{\sqrt{14}\pi m},\textcolor{red}{\mp}\frac{\sqrt{5}}{\sqrt{14}\pi m},\frac{1}{\sqrt{14}\pi m}\right)$&4&No\\
     \hline
     $\pm\left(\frac{3\sqrt{3}}{2\sqrt{13}\pi m},-\frac{\sqrt{3}}{2\sqrt{13}\pi m},0,\textcolor{red}{\pm}\frac{2\sqrt{2}}{\sqrt{13}\pi m},0\right)$&4&No\\
     \hline
     $\pm\left(\frac{\sqrt{17}\textcolor{red}{\pm}\sqrt{51}}{17\pi m},\frac{\sqrt{17}\textcolor{red}{\mp}\sqrt{51}}{17\pi m},\textcolor{blue}{\pm}\frac{2\sqrt{2(2\textcolor{red}{\mp}\sqrt{3})}}{\sqrt{51}\pi m},\textcolor{blue}{\mp}\frac{2\sqrt{102(2\textcolor{red}{\mp}\sqrt{3})}}{51\pi m}(2\textcolor{red}{\pm}\sqrt{3}),\frac{1}{\sqrt{51}\pi m}\right)$&8&No\\
     \hline
    \end{tabular}
    \caption{$H=S_3\times S_3$: fixed points}
\end{table}

$\bullet$ \underline{$H=S_4$}\\

This case is again very similar to $H=S_4\times \mathbb{Z}_2$, leaving only $\{1,2,3,4\}$ invariant. There are consequently 7 invariant operators:

\begin{equation}
    \begin{aligned}
        &\eps\equiv\sum_{i=1,4}\eps_i\,,\quad \eps_5\,,\quad \eps_6\,,\quad \sigma \equiv \sigma_{1234}\,,\\
        &\sigma_5\equiv\sum_{1\leq i,j,k \leq 4}\sigma_{ijk5}\,,\quad \sigma_6\equiv\sum_{1\leq i,j,k \leq 4}\sigma_{ijk6}\,,\quad \sigma_{56}\equiv\sum_{1\leq i,j\leq 4}\sigma_{ij56}\,.
    \end{aligned}
\end{equation}

There are a total of 66 real fixed points. However 6 are decoupled, 22 other points have $H^+ = S_4\times \mathbb{Z}_2$, 10 have $H^+=S_5$, and 2 have $H^+=S_6$. This leaves 26 interacting fixed points with $H^+=S_4$. Among those exist 2 fixed point with content similar to the $S_6$ interacting solutions, with some flipped signs. There are also 4 fixed points corresponding to tensor products of a one $\mathcal{M}_m\to\mathcal{M}_{m-1}$ flows, one trivial solution and a $N=4,G=S_4$ interacting solution. 8 other solutions look like a tensor product of 1 $\mathcal{M}_m\to\mathcal{M}_{m-1}$ flow and an $N=5,G=S_5$ interacting fixed point, but with the wrong sign. The remaining 20 solutions have almost identical couplings to $H=S_5$ or $H=S_4\times \mathbb{Z}_2$ interacting solutions, up to some signs.

\begin{table}[h]
    \centering
    \footnotesize
    \begin{tabular}{|c|c|c|}
    \hline
       $\left(h_{\epsilon}^*,h_{\epsilon_5}^*,h_{\epsilon_6}^*,g_{\sigma}^*,g_{\sigma_5}^*,g_{\sigma_6}^*,g_{\sigma_{56}}^*\right)$ & \# & Tensor Product\\
     \hline
     $\pm\left(\frac{7}{2\sqrt{55}\pi m},\frac{7}{2\sqrt{55}\pi m},-\frac{\sqrt{5}}{2\sqrt{11}\pi m},-\frac{4}{\sqrt{165}\pi m},\frac{4}{\sqrt{165}\pi m},\textcolor{blue}{\pm}\frac{2\sqrt{7}}{\sqrt{165}\pi m},\textcolor{blue}{\mp}\frac{2\sqrt{7}}{\sqrt{165}\pi m}\right)$&4&No\\
     \hline
     $\pm\left(h_{\epsilon^1}(\alpha_i),h_{\epsilon^2}(\alpha_i),h_{\epsilon^2}(\alpha_i),g_{\sigma_1}(\alpha_i),\textcolor{red}{\pm}g_{\sigma_2}(\alpha_i),\textcolor{red}{\mp}g_{\sigma_2}(\alpha_i),g_{\sigma_3}(\alpha_i)\right)$&12&No\\
     \hline
     $\pm\left(\frac{1}{2\sqrt{2}\pi m},\frac{1}{2\sqrt{2}\pi m},\frac{1}{2\sqrt{2}\pi m},-\frac{1}{\sqrt{6}\pi m},\frac{1}{\sqrt{6}\pi m},-\frac{1}{\sqrt{6}\pi m},\frac{1}{\sqrt{6}\pi m}\right)$&2&No\\
     \hline
     $\pm\left(0,\frac{\sqrt{3}}{2\pi m},-\frac{\sqrt{3}}{2\pi m},\textcolor{red}{\pm}\frac{\sqrt{2}}{\pi m},0,0,0\right)$&4&Yes\\
      \hline
     $\pm\left(0,\frac{\sqrt{3}}{2\pi m},0,\textcolor{red}{\pm}\frac{\sqrt{2}}{\pi m},0,\textcolor{red}{\mp}\frac{\sqrt{2}}{\pi m},0\right)$&4&No\\
      \hline
    \end{tabular}
    \caption{$H=S_4$: fixed points}
\end{table}

$\bullet$ \underline{$H=D_{10}$}\\

$D_{10}=\mathbb{Z}_5\rtimes \mathbb{Z}_2$ is the dihedral group of the regular pentagon, and has order 10. It can be generated with 
the 5-cycle $a=(1\,2\,3\,4\,5)$ and the 2-cycle $b=(1\,2)(3\,5)$.
Despite its small size, only 5 invariant operators can be built:
\begin{equation}
    \begin{aligned}
        &\eps=\sum_{i=1}^5\eps_i\,,\quad \eps_6\,,\quad\sigma_1 = \sigma_{1234}+\sigma_{1345}+\sigma_{1245}+\sigma_{2345}+\sigma_{1235}\,,\\
        &\sigma_2=\sigma_{1236}+\sigma_{1456}+\sigma_{2346}+\sigma_{3456}\,,\\
        &\sigma_3=\sigma_{1356}+\sigma_{1246}+\sigma_{2356}+\sigma_{2456}\,.\\
    \end{aligned}
\end{equation}

There are 18 real fixed points, among which only 2 are interacting with $H^+=D_{10}$.

\begin{table}[h]
    \centering
    \footnotesize
    \begin{tabular}{|c|c|c|}
    \hline
       $\left(h_{\epsilon}^*,h_{\epsilon_6}^*,g_{\sigma_1}^*,g_{\sigma_2}^*,g_{\sigma_3}^*\right)$ & \# & Tensor Product\\
     \hline
     $\pm\left(\frac{\sqrt{3}}{2\sqrt{46}\pi m},\frac{\sqrt{3}}{2\sqrt{46}\pi m},-\frac{3}{\sqrt{46}\pi m},-\frac{3}{\sqrt{46}\pi m},\frac{3}{\sqrt{46}\pi m}\right)$&2&No\\
      \hline
    \end{tabular}
    \caption{$H=D_{10}$: fixed points}
\end{table}

\begin{center}
\textbf{Appendix C: invariant couplings and fixed points for $PSL_2(q)$, $q=7,11,13$}    
\end{center}

Finite (simple) groups of Lie type can be subgroups of $S_N$ starting from $N=7$ \cite{Note2}. Unfortunately only a few of them are small enough to be studied with our direct approach. These are $PSL_2(q)$ with $q=7,11,13$. We now describe in detail the cases $N=7,11$. \\

$\bullet$ \underline{$H=PSL_2(7)\cong PSL_3(2),\,N=7$}\\

While generally $PSL_2(q)\subset S_{q+1}$, $q=7,11$ are exceptions in the sense that they are also subgroups of $S_q$. In particular, $PSL_2(7)$ can be generated with the cycles $a=(3\,4)(5\,6)$ and $b=(1\,2\,3)(4\,5\,7)$. This only leaves 3 invariant operators:
\begin{equation}
    \begin{aligned}
        \eps=\sum_{i=1}^7\eps_i\,,\quad \sigma_1=\sigma_{1236}+\sigma_{1245}+\sigma_{1347}+\sigma_{1567}+\sigma_{2357}+\sigma_{2467}+\sigma_{3456}\,,\quad\sigma_2=\sum_{1\leq i,j,k,l\leq 7}\sigma_{ijkl}-\sigma_1\,.
    \end{aligned}
\end{equation}

There are only 6 fixed points, among which 4 are the $S_7$ fixed points. This leaves only two interacting fixed points: 
\begin{table}[h]
    \centering
    \footnotesize
    \begin{tabular}{|c|c|c|}
    \hline
       $\left(h_{\epsilon}^*,g_{\sigma_1}^*,g_{\sigma_2}^*\right)$ & Degeneracy & Tensor Product\\
     \hline
     $\pm\left(\frac{1}{2\pi m},-\frac{1}{\sqrt{3}\pi m},0\right)$&2&No\\
      \hline
    \end{tabular}
    \caption{$H=PSL_2(7)\subset S_7$: fixed points}
\end{table}

$PSL_2(7)$ is maximal in $A_7$, which is maximal in $S_7$. Since the fixed points of $A_7$ are those of $S_7$, we know that the two fixed points we just found are not invariant under any larger subgroup strictly containing $H$.\\

$\bullet$ \underline{$H=PSL_2(11),\,N=11$}\\

$PSL_2(11)$ can be generated with the cycles $a=(1\,5\,11\,9\,10)(2\,7\,6\,3\,4)$ and $b=(1\,2\,8)(3\,7\,9)(5\,6\,10)$. This leaves 4 invariant operators, whose expression are too cumbersome to put even in an appendix. They will be given in the accompanying \texttt{Mathematica} notebook.\\

There are 8 real fixed points. Excluding the 4 $S_{11}$ fixed points, this leaves 4 interacting solutions.

\begin{table}[h]
    \centering
    \footnotesize
    \begin{tabular}{|c|c|c|}
    \hline
       $\left(h_{\epsilon}^*,g_{\sigma_1}^*,g_{\sigma_2}^*,g_{\sigma_2}^*\right)$ & Degeneracy & Tensor Product\\
     \hline
     $\pm\left(\frac{9}{2\sqrt{47}\pi m},\frac{1}{\sqrt{141}\pi m},-\frac{1}{\sqrt{141}\pi m},-\frac{1}{\sqrt{141}\pi m}\right)$&2&No\\
      \hline
     $\pm\left(h_{\eps}(\alpha),g_{\sigma_1}(\alpha),g_{\sigma_2}(\alpha),g_{\sigma_3}(\alpha)\right)$&2&No\\
      \hline
    \end{tabular}
    \caption{$H=PSL_2(11)\subset S_{11}$: fixed points}
\end{table}

where $h_{\epsilon}(x),g_{\sigma_1}(x),g_{\sigma_2}(x),g_{\sigma_3}(x)$ are degree 13 polynomials whose coefficient are given in the \texttt{Mathematica} notebook, and $\alpha\simeq0.0048212$ is the first root of the degree 5 polynomial $P_5(x)= 17692998829316469x^5-332639021682324x^4+3271944017907x^3-14013116172x^2+62354128x-207936$. Numerically these evaluate to
\begin{equation}
    (h_{\eps}(\alpha),g_{\sigma_1}(\alpha),g_{\sigma_2}(\alpha),g_{\sigma_3}(\alpha))\simeq \frac{1}{\pi m}(0.661147,0.109747,-0.0694349,-0.0562615) \,.
\end{equation}

$PSL_2(11)$ is maximal in $M_{11}$, which is maximal in $A_{11}$, itself being maximal in $S_{11}$. Since the fixed points of $M_{11}$ are those of $S_{11}$, we know that the two fixed points we just found are not invariant under any larger subgroup strictly containing $H$.\\

\begin{center}
\textbf{Appendix D: $S_{N}\times S_M$ fixed points for general $N,M$}    
\end{center}

There many infinite families of subgroups that one can study. In order to have hope that general results can be obtained, we need a family of subgroups such that the number of invariant doesn't grow with $N$. A simple example of such a family consists of $H=S_{N}\times S_M\subset S_{N+M}$. In this case, there are always 7 invariant operators:

\begin{equation}
    \begin{aligned}
        &\eps^N=\sum_{i=1}^N\eps_i\,,\quad \eps^M = \sum_{i=N+1}^{N+M}\eps_i\,,\quad \sigma^N = \sum_{1\leq i,j,k,l\leq N}\sigma_{ijkl}\,,\quad \sigma^M = \sum_{N+1\leq i,j,k,l\leq N+M}\sigma_{ijkl}\,,\\
        &\tilde{\sigma}^N = \sum_{1\leq i,j,k\leq N}\sum_{l=N+1}^{N+M}\sigma_{ijkl}\,,\quad \tilde{\sigma}^M = \sum_{i=1}^{N}\sum_{N+1\leq j,k,l\leq N+M}\sigma_{ijkl}\,,\quad \sigma^{MN}=\sum_{1\leq i,j\leq N}\sum_{N+1\leq k,l\leq N+M}\sigma_{ijkl}\,.
    \end{aligned}
\end{equation}
The 1-loop $\beta$ function equations read 
\begin{equation}
\label{eq:beta}
    \begin{aligned}
        &h_{\eps^N}^2+\frac{3}{8}\left(\binom{N-1}{3}g_{\sigma^N}^2+M\binom{N-1}{2}g_{\tilde{\sigma}^N}^2+\binom{M}{3}g_{\tilde{\sigma}^M}^2+(N-1)\binom{M}{2}g_{\sigma^{MN}}^2\right)=\frac{3}{4\pi^2m^2}\,,\\
        &h_{\eps^M}^2+\frac{3}{8}\left(\binom{M-1}{3}g_{\sigma^M}^2+N\binom{M-1}{2}g_{\tilde{\sigma}^M}^2+\binom{N}{3}g_{\tilde{\sigma}^N}^2+(M-1)\binom{N}{2}g_{\sigma^{MN}}^2\right)=\frac{3}{4\pi^2m^2}\,,\\
        &2\sqrt{3}h_{\eps^N}g_{\sigma^N}+3\left(\binom{N-4}{2}g_{\sigma^N}^2+M(N-4)g_{\tilde{\sigma}^N}^2+\binom{M}{2}g_{\sigma^{MN}}^2\right)=0 \,,\\
        &2\sqrt{3}h_{\eps^M}g_{\sigma^M}+3\left(\binom{M-4}{2}g_{\sigma^M}^2+N(M-4)g_{\tilde{\sigma}^M}^2+\binom{N}{2}g_{\sigma^{MN}}^2\right)=0\,,\\
        &\sqrt{3}(3h_{\eps^N}+h_{\eps^M})g_{\tilde{\sigma}^M}+6\left(\binom{N-3}{2}g_{\sigma^N}g_{\tilde{\sigma}^N}+\binom{M-1}{2}g_{\tilde{\sigma}^M}g_{\sigma^{MN}}+(N-3)(M-1)g_{\sigma^{MN}}g_{\tilde{\sigma}^N}\right)=0\,,\\
        &\sqrt{3}(3h_{\eps^M}+h_{\eps^N})g_{\tilde{\sigma}^N}+6\left(\binom{M-3}{2}g_{\sigma^M}g_{\tilde{\sigma}^M}+\binom{N-1}{2}g_{\tilde{\sigma}^N}g_{\sigma^{MN}}+(M-3)(N-1)g_{\sigma^{MN}}g_{\tilde{\sigma}^M}\right)=0\,,\\
        &\sqrt{3}(2h_{\eps^N}+2h_{\eps^M})g_{\sigma^{MN}}+2\Big((N-2)(M-2)g_{\tilde{\sigma}^N}g_{\tilde{\sigma}^M}+\binom{N-2}{2}g_{\sigma^{MN}}g_{\sigma^N}\\
        &+\binom{M-2}{2}g_{\sigma^{MN}}g_{\sigma^M}+2\binom{N-2}{2}g_{\tilde{\sigma}^N}^2+2\binom{M-2}{2}g_{\tilde{\sigma}^M}^2\Big)=0\,.\\
    \end{aligned}
\end{equation}

We assumed that $N,M\geq 4$, meaning $N+M\geq 8$. If $N=3$, then $\sigma^N$ does not exist. If $N=M=3$, $\sigma^N$ and $\sigma^M$ don't exist, and find ourselves back to the $S_3\times S_3 \subset S_6$ scenario with 5 couplings that we studied in (\ref{eq:s3s3}).\\

\begin{figure}
    \centering
\includegraphics[scale=0.5]{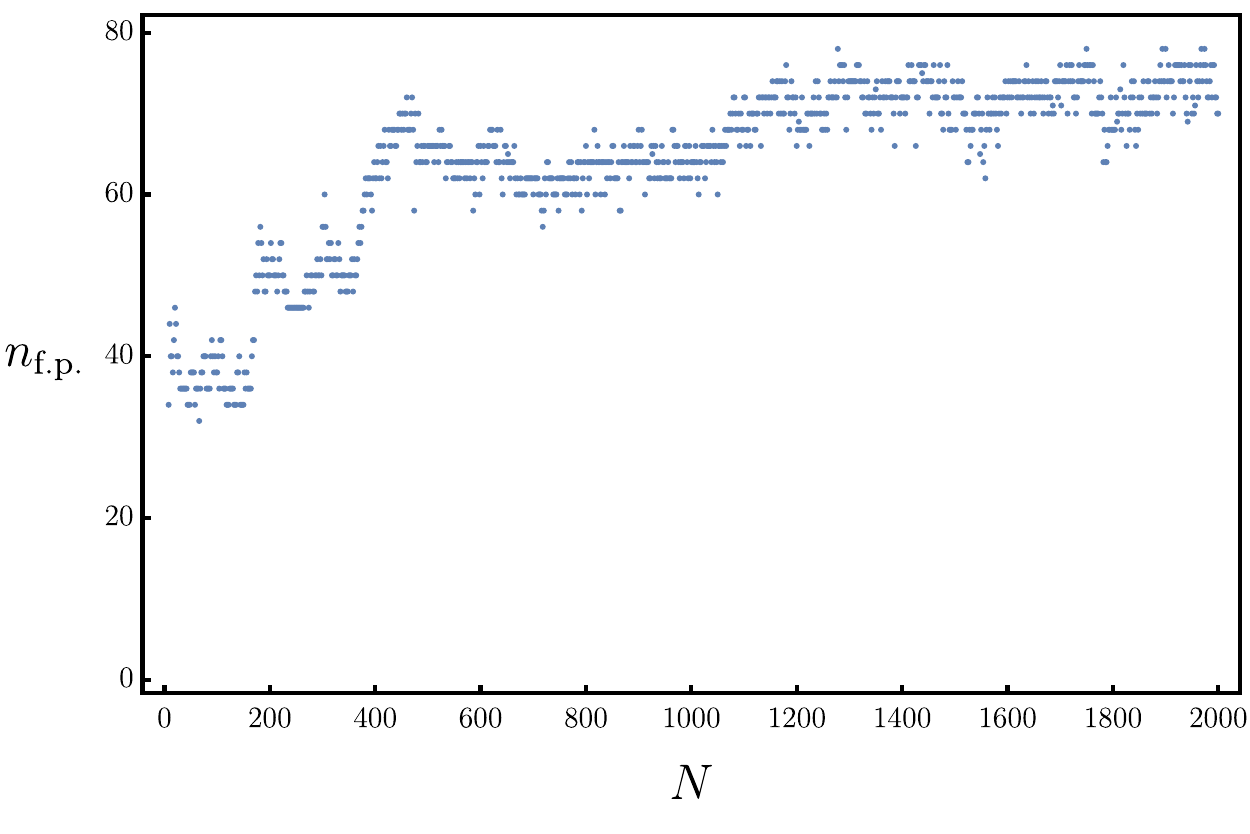}
    \caption{Number of real fixed points invariant under $H=S_{N/2}\times S_{N/2}\subset S_N$.}
\label{fig:snsmn}
\end{figure}

We were not able to solve this system of equations in general but we can still study the number of real fixed points by solving the system of equations (\ref{eq:beta}) numerically for given $N$ and $M$, see Figure (\ref{fig:snsmn}). 

We can simplify our task by not only imposing $N=M$, but additionally imposing a $\mathbb{Z}_2$ symmetry that exchanges the two $S_N$, resulting in $H=(S_{N}\times S_{N})\rtimes \mathbb{Z}_2\subset S_{2N}$. In that case, there are only four couplings, because $h_{\eps^N}=h_{\eps^M}$, $g_{\sigma^N} = g_{\sigma^M}$ and $g_{\tilde\sigma^N} = g_{\tilde\sigma^M}$. There are 16 solutions, 4 being the $S_{2N}$ fixed points. Among the 12 remaining solutions 2 fixed points are the tensor product of two identical interacting $S_{N/2}$ solutions, while the rest are genuinely new. The fixed points are presented in Table X

\begin{table}[h]
    \centering
    \footnotesize
    \begin{tabular}{|c|c|c|}
    \hline
       $\left(h_{\epsilon}^*,g_{\sigma}^*,g_{\tilde\sigma}^*,g_{\sigma^{MN}}^*\right)$ & Degeneracy & Tensor Product\\
     \hline
     $\pm\left(\frac{Q(N)}{2\sqrt{P(N)}\pi m},-\frac{2\sqrt{3}}{\sqrt{P(N)}\pi m},0,0\right)$&2&Yes\\
      \hline
     $\pm\left(\frac{Q(2N)}{2\sqrt{P(2N)}\pi m},-\frac{2\sqrt{3}}{\sqrt{P(2N)}\pi m},\frac{2\sqrt{3}}{\sqrt{P(2N)}\pi m},-\frac{2\sqrt{3}}{\sqrt{P(2N)}\pi m}\right)$&2&No\\
      \hline
     $\pm\left(\frac{3(N-4)R_1(N)}{2\sqrt{R_0(N)}\pi m},-\frac{2\sqrt{3}N(N-3)R_2(N)}{\sqrt{R_0(N)}\pi m},\pm\frac{\sqrt{6(N-4)R_3(N)}}{\sqrt{R_0(N)}\pi m},\frac{2\sqrt{3}(N-2)(N-4)(3N-8)}{\sqrt{R_0(N)}\pi m}\right)$&4&No\\
      \hline
     $\pm\left(\frac{(r_4(N)\textcolor{red}{\pm}\sqrt{r_1(N)})\sqrt{3r_3(N)}}{\sqrt{2(r_2(N)\textcolor{red}{\pm}2\sqrt{r_1(N)})}r_5(N)\pi m},-\frac{3(r_6(N)\textcolor{red}{\pm}\sqrt{r_1(N)})\sqrt{2r_3(N)}}{\sqrt{(r_2(N)\textcolor{red}{\pm}2\sqrt{r_1(N)})}r_7(N)\pi m},0,-\frac{3\sqrt{2r_3(N)}}{\sqrt{r_2(N)\textcolor{red}{\pm}2\sqrt{r_1(N)}}\pi m}\right)$&4&No\\
      \hline
    \end{tabular}
    \caption{$H=(S_{N}\times S_{N})\rtimes \mathbb{Z}_2\subset S_{2N}$: fixed points}
\end{table}
where
\begin{equation}
\begin{aligned}
    &P(N) = 3N^4-53N^3+357N^2-1069N+1194\,,\\
    &Q(N)=3N^2-27N+60\,,\\
&R_0(N)=48N^{12}-1520N^{11}+21592N^{10}-180984N^9+992851N^8-3743459N^7+9925023N^6-18627577N^5\\
&+24591962N^4-22365792N^3+13401728N^2-4796416N+786432\,,\\
&R_1(N)=4N^5-48N^4+209N^3-409N^2+360N-128\,,\\
&R_2(N)=4N^2-15N+8\,,\\
&R_3(N)=-4N^6+86N^5-627N^4+2081N^3-3288N^2+2240N-512\,,\\
&r_1(N)=(N-3)^2(N-2)^6(8N^2-85N+213)^2(N^4+22N^3-99N^2+16N+64)\,,\\
&r_2(N)=34N^8-924N^7+11214N^6-75997N^5+311433N^4-791883N^3+1222363N^2-1043664N+376704\,,\\
&r_3(N)=(N^2-17N+48)^2\,,\\
&r_4(N)=32N^8-948N^7+11760N^6-79832N^5+325020N^4-814944N^3+1233056N^2-1032960N+368064\,,\\
&r_5(N)=(N-2)^2(N^2-17N+48)(8N^2-85N+213)\,,\\
&r_6(N)=16N^8-378N^7+3756N^6-20638N^5+68940N^4-143952N^3+184096N^2-132192N+40896\,,\\
&r_7(N)=(N-3)(N-2)^3(N^2-17N+48)(8N^2-85N+213)\,.
\end{aligned}
\end{equation}
Note that while $R_0(N),r_1(N),r_3(N),r_2-2\sqrt{r_1(N)}>0$ for all $N\geq4$, $R_3(N)$ is positive only for $4\leq N\leq10$. Therefore, there are 16 real solutions only for $4\leq N\leq10$, and only 12 beyond.\\

\begin{center}
\textbf{Appendix E: $(\mathbb{Z}_{2})^{N/2}\rtimes S_{N/2}$ fixed points for general $N$}    
\end{center}

Another subgroup that can be studied for generic $N$ is $H=(\mathbb{Z}_{2})^{N/2}\rtimes S_{N/2}$. Only 4 invariant operators can be constructed, allowing us to solve the beta function equations at leading order for arbitrary $N$:
\begin{equation}
    \begin{aligned}
        &\eps=\sum_{i=1}^{N}\eps_{i}\,,\quad \sigma_0 = \sum_{\eta_{1},\eta_2\eta_{3},\eta_4=0,1}\sum_{1\leq i<j<k<l\leq N/2}\sigma_{2i-\eta_1,2j-\eta_2,2k-\eta_2,2l-\eta_4}\,,\\&\sigma_2 = \sum_{\eta_{3},\eta_4=0,1}\sum_{1\leq i<j<k\leq N/2}\sigma_{2i-1,2i,2j-\eta_3,2k-\eta_4}\,,\quad
      \sigma_4 = \sum_{1\leq i<j\leq N/2}\sigma_{2i-1,2i,2j-1,2j}\,.
    \end{aligned}
\end{equation}
    
For $N\geq 10$, only 10 real fixed point exist, including the 4 $S_N$ fixed points. The exact expressions of 4 of the remaining fixed points can be worked out exactly, but is too cumbersome to be written down here. As for the other two, they read as:
\begin{equation}
    \begin{aligned}
        \left(h_{\epsilon}^*,g_{\sigma_1}^*,g_{\sigma_2}^*,g_{\sigma_4}^*\right) = \pm\left(\frac{\sqrt{3}}{2\pi m}\frac{N-4}{\sqrt{N^2+N-2}},0,0,-\frac{6}{\pi m}\frac{1}{\sqrt{N^2+N-2}}\right)
    \end{aligned}
\end{equation}


\begin{thebibliography}{99}
\bibitem{NoteD12}
The case of $D_{12}$ is harder to classify because different classes of fixed points exist depending on how we embed the subgroup in $S_6$. Nonetheless, we do find non-trivial fixed points with $D_{12}$ symmetry.
\end{thebibliography}
\end{document}